\documentclass[preprintnumbers,showpacs,amsmath,amssymb,floatfix,prd,onecolumn,superscriptaddress,nofootinbib,12pt]{revtex4}
\usepackage{amssymb}
\usepackage{graphicx}
\usepackage{epsfig}
\usepackage{bm}
\usepackage{mathrsfs}
\usepackage{amsfonts}
\usepackage{epstopdf}
\usepackage{bm}
\usepackage{amsthm}
\usepackage{multirow}
\usepackage{subfigure}
\usepackage{amsmath}
\usepackage{textcomp}
\usepackage{fancyhdr}
\usepackage{tipa}
\usepackage{babel}
\begin{document}
	\title{Off-Shell ADT Conserved Quantities in Palatini Gravity}
	
	\author{Hai-Feng \surname{Ding}}
	\email[]{haifeng1116@qq.com}
	\author{Xiang-Hua \surname{Zhai}}
	\email[]{zhaixh@shnu.edu.cn}
	\affiliation{Division of Mathematical and Theoretical Physics, Shanghai Normal University, 100 Guilin Road, Shanghai 200234, China}

\begin{abstract}

In this paper we generalize the off-shell Abbott-Deser-Tekin (ADT) conserved charge formalism to Palatini theory of gravity with torsion and non-metricity. Our construction is based on the coordinate formalism and the independent dynamic fields are the metric and the affine connection. For a general Palatini theory of gravity, which is diffeomorphism invariant up to a boundary term, we obtain the most general expression for off-shell ADT potential. As explicit examples, we derive the off-shell ADT potentials for Einstein-Hilbert action, the most general $L(g_{\mu \nu}, R^{\lambda}{}_{\nu \alpha \mu}, T^{\lambda}{}_{\alpha \beta}, Q_{\alpha \mu \nu})$ theories and the teleparallel Palatini gravity.

\end{abstract}

%\pacs{14.80.bn, 98.80.Es, 98.80.Cq }
\maketitle

%##############################################

\maketitle

%###########################################################################%
\newpage
\section{Introduction}

Symmetries and conservation laws play important roles in modern physics. In general theories of gravity, the construction of conserved quantities is a complex procedure. In order to obtain an effective way for computing the conserved quantities, several approaches have been proposed.

In the well known ADM approach \cite{Arnowitt:1962hi}, the conserved charges could be calculated for asymptotically flat spacetime at the asymptotic infinity. The extension to asymptotic (anti-)de sitter geometry has been given by Abbott-Deser-Tekin (ADT) \cite{Abbott:1981ff,Abbott:1982jh,Deser:2002rt,Deser:2002jk,Senturk:2012yi} in a covariant manner. Based on the Noether’s theorem, the covariant phase space method (CPSM) was proposed by Wald and Iyer \cite{Wald:1993nt,Iyer:1994ys,Wald:1999wa}. In the framework of CPSM, a convenient method was developed for calculating conserved charges associated with ‘exact symmetry’ of black hole solutions  \cite{Hajian:2015xlp,Hajian:2016kxx,Ghodrati:2016vvf}, which is called solution phase space method (SPSM). To define the quasi-local energy, a covariant Hamiltonian method was developed by Chen, Nester and Tung \cite{Chen:1994qg,Chen:1998aw,Chen:2005hwa,Chen:2015vya}. Along another line, general theory of conserved charges based on the cohomology principles was developed by Barnich-Brandt-Comp\`ere (BBC) \cite{Barnich:2001jy,Barnich:2003xg,Barnich:2004uw,Barnich:2007bf}. Recently, an off-shell ADT formalism was proposed by Bouchareb, Cl\'ement\cite{Bouchareb:2007yx} and Kim-Kulkami-Yi in \cite{Kim:2013zha}, and subsequently was generalized to including the matter fields \cite{Hyun:2014sha,Ding:2019dhy}. 

Each of these methods has its own merits and demerits, and most of these works were confined to pure Riemannian geometrical framework of gravity theories. Extentions to non-Riemannian geometries were studied in \cite{Hehl:1994ue,Julia:1998ys,Julia:2000er,Giachetta:1995bj,Vollick:2007fh}, and recently in \cite{Adami:2017phg,Obukhov:2015eha,Chakraborty:2018qew,Emtsova:2019moq,Hammad:2019oyb,Barnich:2020ciy}. The main property of non-Riemannian geometries is that the torsion and the non-metricity are included. The most general theory in non-Riemannian geometries is the Palatini gravity, which is called the Metric-Affine-Gravity (MAG) \cite{Hehl:1994ue,Iosifidis:2019jgi} when the matter fields are included. Usually, the MAG is expressed in terms of two formalisms. The first one is written in the language of exterior differential forms, and the independent dynamic fields are the co-frame $e^a$ and the linear connection $\omega^a{}_b$, thus it is also named the frame-vielbein formalism. The second one is the coordinate formalism (namely, the usual tensoral language), and the independent dynamic fields are the metric $g_{\mu \nu}$ and the affine connection $\Gamma^{\lambda}{}_{\mu \nu}$.

To our knowledge, till now most of the constructions of conserved quantities in the non-Riemannian geometrical framework are in the language of exterior differential forms and are on shell, i.e. the equations of motion (EOM) are used. On the other hand, as mentioned in \cite{Chakraborty:2018qew}, when torsion is present, the derivations of the off-shell conserved quantities are complicated. The main reason is that the diffeomorphism variations of the metric and the affine connection cannot be written in terms of derivatives of the diffeomorphism vector fields. We will attempt to solve this problem by using the off-shell ADT formalism. In this paper, we will generalize the off-shell ADT formalism to the Palatini theory of gravity with torsion and non-metricity in the coordinate formalism.

The paper is organized as follows. In Sect. 2, some basic formulas in Palatini theory are presented for our subsequent derivation. In Sect. 3, we generalize the off-shell ADT formalism to Palatini theory with torsion and non-metricity. In Sect. 4, we derive the off-shell ADT potentials for Einstein-Hilbert action, the most general $L(g_{\mu \nu}, R^{\lambda}{}_{\nu \alpha \mu}, T^{\lambda}{}_{\alpha \beta}, Q_{\alpha \mu \nu})$ theories and the teleparallel Palatini gravity. In Sect. 5, the conserved charges are analyzed if the black hole solutions exist. The conclusions are given in the last section.

%###########################################################################%
\section{Palatini theories}

We first fix our notations and conventions by briefly reviewing the basic geometrical objects in Palatini theory with torsion and non-metricity in the coordinate formalism \cite{Iosifidis:2019jgi}, which will be used in our work.

We define the affine connection and the associated covariant derivative as 
\begin{align}
   \nabla_{\mu}u^{\nu}=\partial_{\mu}u^{\nu}+\Gamma^{\nu}{}_{\mu \sigma}u^{\sigma} ,
\end{align}
where $u^{\nu}$ is an arbitrary vector field. In order to perform our subsequent derivation and make the expressions compact we define the transpose connection as $\bar \Gamma^{\lambda}{}_{\mu \nu}=\Gamma^{\lambda}{}_{\nu \mu}$, and the associated covariant derivative as $\bar \nabla_{\mu}$,
\begin{align}
	\bar\nabla_{\mu}u^{\nu}=\partial_{\mu}u^{\nu}+\bar\Gamma^{\nu}{}_{\mu \sigma}u^{\sigma}=\partial_{\mu}u^{\nu}+\Gamma^{\nu}{}_{\sigma \mu}u^{\sigma} .
\end{align}
The basic geometrical objects in Palatini theory are torsion, non-metricity and curvature, which are defined as follows
\begin{align}\label{gobjects}
   T^{\lambda}{}_{\mu \nu}=& -2\Gamma^\lambda{}_{[\mu \nu]},  \nonumber \\
   Q_{\alpha \mu \nu}=& -\nabla_{\alpha} g_{\mu \nu}=-\partial_{\alpha} g_{\mu \nu}+\Gamma^{\sigma}{}_{\alpha \mu} g_{\sigma \nu}+\Gamma^{\sigma}{}_{\alpha \nu} g_{\mu \sigma},  \nonumber \\
   R^{\mu}{}_{\nu \alpha \beta}=& 2\partial_{[\alpha} \Gamma^{\mu}{}_{\beta] \nu}+2\Gamma^{\mu}{}_{[\alpha | \sigma|} \Gamma^{\sigma}{}_{\beta ] \nu} .
\end{align}
The commutators of covariant derivatives acting on a scalar $\phi$ and a vector $u^{\mu}$ are given by
\begin{align}
   \left[\nabla_{\alpha}, \nabla_{\beta}\right] \phi=& T^{\lambda}{}_{\alpha \beta} \nabla_{\lambda} \phi,  \nonumber \\
   \left[\nabla_{\alpha}, \nabla_{\beta}\right] u^{\mu}=& R^{\mu}{}_{\nu \alpha \beta} u^{\nu}+T^{\lambda}{}_{\alpha \beta} \nabla_{\lambda} u^{\mu} .
\end{align}
It is easy to generalize the covariant derivatives and the commutators acting on arbitrary rank tensors. The contractions of the basic geometrical objects have the forms
\begin{align}
R_{\nu \beta}=&R^{\mu}{}_{\nu \mu \beta},  &R&=g^{\nu \beta}R_{\nu \beta}, \nonumber \\
T_{\mu}=&T^{\lambda}{}_{\mu \lambda},  &Q_{\alpha}&=g^{\mu \nu} Q_{\alpha \mu \nu}=Q_{\alpha \mu}{}^{\mu} ,  \nonumber \\ \check{R}^{\mu}{}_{\beta}=&g^{\nu \alpha} R^{\mu}{}_{\nu \alpha \beta},   &\hat {R}_{\alpha \beta}&=R^{\mu}{}_{\mu \alpha \beta}=\partial_{[\alpha} Q_{\beta]} ,
\end{align}
where $R_{\nu \beta}$ is Ricci tensor, which is not symmetric in Palatini theory. The affine connection can be decomposed as
\begin{equation}\label{key}
   \Gamma^{\lambda}{}_{\mu \nu}=\mathring\Gamma^{\lambda}{}_{\mu \nu}+N^{\lambda}{}_{\mu \nu},
\end{equation}
where $\mathring\Gamma^{\lambda}{}_{\mu \nu}$ and $N^{\lambda}{}_{\mu \nu}$ are Levi-Civita connection and the distorsion tensor, respectively, which are given by 
\begin{align}
   \mathring\Gamma^{\lambda}{}_{\mu \nu}=&\frac{1}{2} g^{\lambda \alpha}(\partial_{\mu}g_{\nu \alpha}+\partial_{\nu} g_{\mu \alpha}-\partial_{\alpha} g_{\mu \nu}),  \nonumber \\
   N^{\lambda}{}_{\mu \nu}=&K^{\lambda}{}_{\mu \nu}+L^{\lambda}{}_{\mu \nu}  \nonumber \\
   =&\frac{1}{2}g^{\lambda \alpha}(T_{\mu \nu \alpha}+T_{\nu \mu \alpha}-T_{\alpha \mu \nu})+\frac{1}{2}g^{\lambda \alpha}(Q_{\mu \nu \alpha}+Q_{\nu \mu \alpha}-Q_{\alpha \mu \nu}),
\end{align}
where $L^{\lambda}{}_{\mu \nu}$ and $K^{\lambda}{}_{\mu \nu}$ are called the disformation and the contorsion tensors, respectively.

Using the definition of the Riemann tensor and applying the Jacobi identity of covariant derivatives to scalars and vectors, one can obtain the Bianchi identities 
\begin{align}
   R^{\mu}{}_{\nu (\alpha \beta)}=&0, \\
   R^{\lambda}{}_{[\alpha \beta \gamma]}+\nabla_{[\alpha} T^{\lambda}{}_{\beta \gamma]}+T^{\sigma}{}_{[\alpha \beta} T^{\lambda}{}_{\gamma] \sigma}=&0, \label{BI8}  \\
   \nabla_{[\alpha} R^{\sigma}{}_{|\rho| \beta \gamma]}+T^{\lambda}{}_{[\alpha \beta} R^{\sigma}{}_{|\rho| \gamma] \lambda}=&0. \label{BI9}
\end{align}
The variations of the basic geometrical objects have the following forms
\begin{align}
   \delta_g T^{\lambda}{}_{\mu \nu}=&0, &\delta_{\Gamma}T^{\lambda}{}_{\alpha \beta}&=-2\delta^{[\mu}{}_{\alpha}\delta^{\nu]}{}_{\beta} \delta \Gamma^{\lambda}{}_{\mu \nu}, \nonumber \\
   \delta_g R^{\mu}{}_{\nu \alpha \beta}=&0,  &\delta_{\Gamma}R^{\mu}{}_{\nu \alpha \beta}&=\nabla_{\alpha}\delta\Gamma^{\mu}{}_{\beta \nu}-\nabla_{\beta}\delta\Gamma^{\mu}{}_{\alpha \nu}-T^{\lambda}{}_{\alpha \beta}\delta\Gamma^{\mu}{}_{\lambda \nu}, \nonumber \\
   \delta_g Q_{\rho \mu \nu}=&-\nabla_{\rho} \delta g_{\mu\nu}, &\delta_{\Gamma} Q_{\rho \alpha \beta}&=2 \delta^{\mu}{}_{\rho} \delta^{\nu}{}_{(\alpha} g_{\beta) \lambda} \delta \Gamma^{\lambda}{}_{\mu \nu},
\end{align}
where the torsion and the curvature tensors are independent of the metric.

The covariant derivatives of metric determinant and the non-trivial surface term are given by
\begin{align}
   \nabla_{\mu}\sqrt{-g}=&\partial_{\mu}\sqrt{-g}-\Gamma^{\alpha}{}_{\mu \alpha}\sqrt{-g},  \nonumber \\
   \nabla_{\mu}(\sqrt{-g} u^{\mu})=&\partial_{\mu}(\sqrt{-g} u^{\mu})+\sqrt{-g}T_{\mu}u^{\mu}, \nonumber \\
   \nabla_{[\mu}Q_{\nu] \alpha \beta}=&R_{(\alpha \beta) \mu \nu}+\frac{1}{2}T^{\lambda}{}_{\mu \nu}Q_{\lambda \alpha \beta}.
\end{align}

\section{Generalized off-shell ADT current and potential in Palatini theory}

In this section we construct the off-shell ADT current and potential in Palatini theory of gravity with torsion and non-metricity. In Palatini formalism, the two independent dynamic fields are the metric $g_{\mu \nu}$ and the affine connection $\Gamma^{\lambda}{}_{\mu \nu}$.

\subsection{Off-shell currents} 

We consider a general Palatini theory of gravity with action  
\begin{equation}\label{key}
   I[g,\Gamma]=\frac{1}{16 \pi G}\int d^Dx \sqrt{-g} L(g, \Gamma).
\end{equation}
In this paper, we suppose that the theories under consideration are diffeomorhism invariant up to a boundary term. For convenience, we denote the dynamic fields jointly as $\Phi=(g_{\mu \nu},\Gamma^{\lambda}{}_{\mu \nu})$. The generic variation of the Lagrangian leads to 
\begin{align} \label{Lvariation}
   \delta (\sqrt{-g} L)=&\sqrt{-g} \mathcal{E}_{\Phi }\delta \Phi+\partial_{\mu} (\sqrt{-g} \Theta^{\mu}(\delta \Phi, \Phi) ) \nonumber \\
   =&-\sqrt{-g}\mathcal{E}^{\mu \nu} \delta g_{\mu \nu}+\sqrt{-g} P_{\lambda}{}^{\nu \mu} \delta \Gamma^{\lambda}{}_{\mu \nu}+\partial_{\mu} \tilde\Theta^{\mu}(\delta \Phi, \Phi),
\end{align}
where $\mathcal{E}_{\Phi }=(\mathcal{E}_{\mu \nu}, P_{\lambda}{}^{\nu \mu} )$ and $\tilde\Theta^{\mu}=\sqrt{-g} \Theta^{\mu}$ denote the Euler-Lagrange expression and the surface term, respectively. And hereinafter, the tilde $\tilde{} $  over a letter denotes $\tilde {X}=\sqrt{-g}X$.

For an arbitrary smooth vector field $\xi=\xi^{\mu} \partial_{\mu}$ defined over the spacetime, which can generate the diffeomorhism $x^{\mu }\to x^{\mu }-\xi ^{\mu }$, the diffeomorphism variation of the Lagrangian is given by 
\begin{align} \label{eq14}
   \delta_{\xi} (\sqrt{-g} L)=& \sqrt{-g} \mathcal{E}_{\Phi} \delta_{\xi} \Phi+\partial_{\mu}\tilde \Theta^{\mu}(\delta_{\xi} \Phi, \Phi)  \nonumber \\
   =&-\sqrt{-g}\mathcal{E}^{\mu \nu} \delta_{\xi} g_{\mu \nu}+\sqrt{-g} P_{\lambda}{}^{\nu \mu} \delta_{\xi} \Gamma^{\lambda}{}_{\mu \nu}+\partial_{\mu} \tilde\Theta^{\mu}(\delta_{\xi} \Phi, \Phi),
\end{align}
where the diffeomorphism transformations have the forms \cite{Obukhov:2015eha}
\begin{align}
   \delta_{\xi} g_{\mu \nu}=&\xi^{\sigma} \partial_{\sigma} g_{\mu \nu}+g_{\sigma \nu} \partial_{\mu}\xi^{\sigma}+g_{\mu \sigma} \partial_{\nu} \xi^{\sigma}  \nonumber \\
   =&2\nabla_{(\mu} \xi_{\nu)}+2N^{\lambda}{}_{(\mu \nu)} \xi_{\lambda},  \nonumber \\
   \delta_{\xi} \Gamma^{\lambda}{}_{\mu \nu}=&\partial_{\mu} \partial_{\nu} \xi^{\lambda}+\xi^{\sigma} \partial_{\sigma} \Gamma^{\lambda}{}_{\mu \nu}-\Gamma^{\sigma}{}_{\mu \nu} \partial_{\sigma} \xi^{\lambda}+\Gamma^{\lambda}{}_{\sigma \nu} \partial_{\mu} \xi^{\sigma}+\Gamma^{\lambda}{}_{\mu \sigma} \partial_{\nu} \xi^{\sigma}  \nonumber \\
   =&\nabla_{\mu} \bar \nabla_{\nu} \xi^{\lambda}-R^{\lambda}{}_{\nu \mu \sigma} \xi^{\sigma}.
\end{align}
Then, one has 
\begin{align} \label{eq16}
   \delta_{\xi} (\sqrt{-g} L)=&\left[2\nabla_{\mu} (\sqrt{-g} \mathcal{E}^{(\mu \nu)})-2 T_{\mu} (\sqrt{-g} \mathcal{E}^{(\mu \nu)})-2\sqrt{-g} \mathcal{E}^{(\mu \lambda)} N^{\nu}{}_{\mu \lambda}-\sqrt{-g} P_{\lambda}{}^{\alpha \beta} R^{\lambda}{}_{\alpha \beta}{}^{\nu} \right] \xi_{\nu}  \nonumber \\
   +&\left[\bar \nabla_{\nu}[\nabla_{\mu}(\sqrt{-g}P_{\sigma}{}^{\nu  \mu})-T_{\mu}(\sqrt{-g}P_{\sigma}{}^{\nu  \mu})]+T_{\nu}[\nabla_{\mu}(\sqrt{-g}P_{\sigma}{}^{\nu  \mu})-T_{\mu}(\sqrt{-g}P_{\sigma}{}^{\nu  \mu})]\right]\xi^{\sigma}  \nonumber \\
   +&\partial_{\mu} (\tilde \Theta^{\mu}(\delta_{\xi} \Phi, \Phi)-2\tilde S^{\mu}(\delta_{\xi} \Phi, \Phi)),
\end{align}
where
\begin{equation}\label{key}
   2\tilde S^{\mu}(\delta_{\xi} \Phi, \Phi) \equiv 2 \sqrt{-g} \mathcal{E}^{(\mu \nu)} \xi_{\nu}-\sqrt{-g} P_{\sigma}{}^{\nu \mu} \bar \nabla_{\nu} \xi^{\sigma}+[\nabla_{\nu}(\sqrt{-g}P_{\sigma}{}^{\mu \nu})-T_{\nu}(\sqrt{-g}P_{\sigma}{}^{\mu \nu})] \xi^{\sigma}.
\end{equation}

On the other hand, the diffeomorphism variation of the Lagrangian has the form 
\begin{equation} \label{eq18}
   \delta_{\xi} (\sqrt{-g} L)=\partial_{\mu}(\xi^{\mu} \sqrt{-g} L).
\end{equation}
By equating the two expressions above, we can obtain an off-shell identity that can be thought of as generalized Bianchi identity in Palatini gravity
\begin{align} \label{eq19}
   &\left[2\nabla_{\mu} (\sqrt{-g} \mathcal{E}^{(\mu \nu)})-2T_{\mu} (\sqrt{-g} \mathcal{E}^{(\mu \nu)})-2\sqrt{-g} \mathcal{E}^{(\mu \lambda)} N^{\nu}{}_{\mu \lambda}-\sqrt{-g} P_{\lambda}{}^{\alpha \beta} R^{\lambda}{}_{\alpha \beta}{}^{\nu} \right] \xi_{\nu}  \nonumber \\
   &+\left[\bar \nabla_{\nu}[\nabla_{\mu}(\sqrt{-g}P_{\sigma}{}^{\nu  \mu})-T_{\mu}(\sqrt{-g}P_{\sigma}{}^{\nu  \mu})]+T_{\nu}[\nabla_{\mu}(\sqrt{-g}P_{\sigma}{}^{\nu  \mu})-T_{\mu}(\sqrt{-g}P_{\sigma}{}^{\nu  \mu})]\right]\xi^{\sigma}  \nonumber \\
   &=\partial_{\mu} (\sqrt{-g} \mathcal{Z}^{\mu}_{\xi}),
\end{align}
where we have defined
\begin{equation}\label{key}
   \sqrt{-g} \mathcal{Z}^{\mu}_{\xi} \equiv \xi^{\mu} \sqrt{-g} L-\tilde \Theta^{\mu}(\delta_{\xi} \Phi, \Phi)+2\tilde S^{\mu}(\delta_{\xi} \Phi, \Phi)+\partial _{\nu }\tilde{U}^{\left[\mu \nu \right]},
\end{equation}
and $\tilde{U}^{\mu \nu }=\tilde{U}^{\left[\mu \nu \right]}$ is an arbitrary anti-symmetric second rank tensor that will be dropped out in what follows since this ambiguity will not affect the final conserved charges.

Furthermore, from Eqs.\eqref{eq14}, \eqref{eq16} and \eqref{eq19} we obtain 
\begin{equation}\label{eq21}
   \partial_{\mu}(2\sqrt{-g} \mathrm{E}^{\mu}_{\xi})=\sqrt{-g} \mathcal{E}^{(\mu \nu)} \delta_{\xi} g_{\mu \nu}-\sqrt{-g} P_{\lambda}{}^{\nu \mu} \delta_{\xi} \Gamma^{\lambda}{}_{\mu \nu}=-\sqrt{-g} \mathcal{E}_{\Phi} \delta_{\xi} \Phi,
\end{equation}
where
\begin{equation}\label{key}
   \mathrm{E}^{\mu}_{\xi} \equiv S^{\mu}_{\xi}-\frac{1}{2} \mathcal{Z}^{\mu}_{\xi}.
\end{equation}
If the transformation $\delta_{\xi} \Phi$ is an exact symmetry $\delta_{\xi} \Phi=0$ \cite{Hajian:2015xlp,Hajian:2016kxx,Ghodrati:2016vvf} (i.e. $\delta_{\xi} g_{\mu \nu}=0$ and $\delta_{\xi} \Gamma^{\lambda}{}_{\mu \nu}=0$, in which the vector field $\xi$ satisfying these two conditions is also called the generalized Killing vector \cite{Obukhov:2015eha}), we can get
\begin{equation}\label{eq23}
   \partial_{\mu}(\sqrt{-g} \mathrm{E}^{\mu}_{\xi})=0.
\end{equation}

To obtain the off-shell conserved current, we consider the double variations \cite{Hyun:2014sha}
\begin{equation}\label{key1}
\delta _1 \delta _2 I\left[\Phi \right] =\frac{1}{16 \pi  G}\int d^Dx \left[\delta _1 \left(\sqrt{-g}\mathcal{E}_{\Phi }\delta _2\Phi\right)+\partial _{\mu }\left(\delta _1\tilde{\Theta }^{\mu }\left(\delta _2 \Phi ,\Phi \right)\right)\right] .
\end{equation}
Using the property
\begin{equation}\label{vproperty}
(\delta _1 \delta _2-\delta _2 \delta _1) I\left[\Phi \right]=0 ,
\end{equation}
in which we take one of the variations as a diffeomorhism variation $\delta _{\xi }$ (In this case, the property \eqref{vproperty} is trivially true, since the action should not vary under a diffeomorhism), we have
\begin{equation}\label{key}
0=\frac{1}{16 \pi  G}\int d^Dx\left[\delta _{\xi } \left(\sqrt{-g} \mathcal{E}_{\Phi }\delta \Phi\right)-\delta  \left(\sqrt{-g}\mathcal{E}_{\Phi }\delta _{\xi }\Phi\right)+\partial_{\mu} \tilde{\omega }^{\mu } \left(\delta \Phi ,\delta _{\xi }\Phi,\Phi \right)\right] ,
\end{equation}
where we have used the symplectic current definition in CPSM \cite{Wald:1993nt,Iyer:1994ys,Wald:1999wa}
\begin{equation}\label{symplectic current}
\tilde{\omega }^{\mu }\left(\delta  \Phi ,\delta _{\xi }\Phi,\Phi \right)=\delta \tilde{\Theta }^{\mu }\left(\delta _{\xi} \Phi,\Phi \right)-\delta _{\xi }\tilde{\Theta }^{\mu }(\delta \Phi ,\Phi ).
\end{equation}
We assume that the spacetime is endowed with a generalized Killing vector i.e. $\delta _{\xi }\Phi  =0$, then $\tilde{\omega }^{\mu }\left(\delta \Phi ,\delta _{\xi }\Phi,\Phi \right)=0$. Since the Euler-Lagrange expression is covariant, we can obtain the off-shell identity\footnote{That is, the identity (29) holds not only on-shell but also off-shell.}
\begin{equation} \label{eq28}
\delta _{\xi } \left(\sqrt{-g} \mathcal{E}_{\Phi }\delta \Phi\right)=\partial_ {\mu} \left(\xi ^{\mu } \sqrt{-g}\mathcal{E}_{\Phi }\delta \Phi\right)=0 . 
\end{equation}

Similar to \cite{Hyun:2014sha,Ding:2019dhy}, we can introduce an off-shell ADT current for the generalized Killing vector 
\begin{equation}\label{off-shell ADT current}
   \sqrt{-g} \mathcal{J}^{\mu}_{\mathrm{ADT}}=\delta(\sqrt{-g} \mathrm{E}^{\mu}_{\xi})+\frac{1}{2} \sqrt{-g} \xi^{\mu} \mathcal{E}_{\Phi} \delta \Phi ,
\end{equation}
where we have taken $\delta\xi ^{\mu }=0$, i.e. the generators are field-independent.
From Eqs.\eqref{eq23} and \eqref{eq28}, it is easy to see the conservation of the off-shell ADT current as
\begin{equation}\label{key}
\partial_ {\mu}\left(\sqrt{-g} \mathcal{J}_{\mathrm{ADT}}^{\mu }\right)=0 .
\end{equation}
So we are allowed to introduce the off-shell ADT potential $\mathcal{Q}_{\mathrm{ADT}}^{\mu \nu }$ as
\begin{equation}\label{key}
   \sqrt{-g}\mathcal{J}_{\mathrm{ADT}}^{\mu }=\partial _{\nu }(\sqrt{-g}\mathcal{Q}_{\mathrm{ADT}}^{\mu \nu}) .
\end{equation}

\subsection{Off-shell potentials}

In this subsection we construct the off-shell ADT potential by using the off-shell Noether current and potential.
From Eqs.\eqref{eq14} and \eqref{eq18}, and using the off-shell identity \eqref{eq21}, we can introduce the off-shell Noether current
\begin{equation}\label{off-shell Noether current}
J_{\xi}^{\mu }=2 \sqrt{-g} \mathrm{E}^{\mu}_{\xi}+\sqrt{-g} \xi ^{\mu }  L-\tilde{\Theta }^{\mu }\left(  \delta _{\xi }\Phi,\Phi \right) ,
\end{equation} 
which satisfies $\partial_{\mu} J_{\xi}^{\mu }$=0. Then, the off-shell Noether potential $K_{\xi }^{\mu \nu }$ can be introduced as 
\begin{equation}\label{Noether current and potentiali}
J_{\xi }^{\mu }=\partial_{\nu} K_{\xi }^{\mu \nu } ,
\end{equation}
where $J_{\xi }^{\mu }=\sqrt{-g} \mathcal{J}_{\xi }^{\mu }$ and $K_{\xi }^{\mu \nu }=\sqrt{-g} \mathcal{K}_{\xi }^{\mu \nu }$.
The diffeomorphism variation of the surface term is
\begin{equation}\label{key}
\delta_{\xi }\tilde{\Theta }^{\mu }(\delta \Phi ,\Phi )=\mathcal{L}_{\xi }\tilde{\Theta }^{\mu }(\delta \Phi ,\Phi )=\xi ^{\nu } \partial_{\nu} \tilde{\Theta }^{\mu }-\tilde{\Theta }^{\nu } \partial_{\nu} \xi ^{\mu }+\tilde{\Theta }^{\mu } \partial_{\nu} \xi ^{\nu } ,
\end{equation}
which leads to
\begin{equation}\label{eq35}
\xi ^{\mu }\partial_{\nu} \tilde{\Theta }^{\nu }(\delta \Phi ,\Phi )=\partial _{\nu }(2 \xi ^{[\mu }\tilde{\Theta }^{\nu ]}(\delta \Phi ,\Phi ))+\mathcal{L}_{\xi }\tilde{\Theta }^{\mu }(\delta \Phi ,\Phi ) .
\end{equation}
By varying the off-shell Noether current \eqref{off-shell Noether current} and using the off-shell ADT current \eqref{off-shell ADT current}, we get
\begin{equation}\label{eq36}
   \delta J^{\mu}_{\xi}=2 \sqrt{-g} \mathcal{J}^{\mu}_{\mathrm{ADT}}+\partial_{\nu}(2 \xi ^{[\mu }\tilde{\Theta }^{\nu ]}(\delta \Phi ,\Phi ))-\tilde \omega^{\mu}(\delta \Phi,\delta_{\xi} \Phi, \Phi),
\end{equation}
where we have used Eq.\eqref{eq35} and the symplectic current definition \eqref{symplectic current}. From Eq.\eqref{eq36} and the variation of \eqref{Noether current and potentiali} $\delta J^{\mu}_{\xi}=\partial_{\nu}(\delta K_{\xi }^{\mu \nu })$ we obtain
\begin{align}
   2 \sqrt{-g} \mathcal{J}^{\mu}_{\mathrm{ADT}} =&\partial_{\nu}(\delta K_{\xi }^{\mu \nu }-2 \xi ^{[\mu }\tilde{\Theta }^{\nu ]}(\delta \Phi ,\Phi ))+\tilde \omega^{\mu}(\delta \Phi,\delta_{\xi} \Phi, \Phi)  \nonumber \\
   =&\partial _{\nu }(2\sqrt{-g}\mathcal{Q}_{\mathrm{ADT}}^{\mu \nu}) ,
\end{align}
where $\delta _{\xi }\Phi  =0$, $\tilde{\omega }^{\mu }\left(\delta \Phi ,\delta _{\xi }\Phi,\Phi \right)=0$, for exact symmetry.
$\mathcal{Q}_{\mathrm{ADT}}^{\mu \nu}$ is the off-shell ADT potential corresponding to the off-shell ADT current $\mathcal{J}_{\mathrm{ADT}}^{\mu }$, which is given by
\begin{align}\label{off-shell ADT potential}
2 \sqrt{-g} \mathcal{Q}_{\mathrm{ADT}}^{\mu \nu}&=\delta K_{\xi }^{\mu \nu }-2\xi ^{[\mu }\tilde{\Theta }^{\nu ]}(\delta \Phi ,\Phi ) .
\end{align}

In order to obtain finite conserved charges, we use the one parameter path integral method \cite{Wald:1999wa,Barnich:2001jy,Barnich:2003xg,Barnich:2004uw,Barnich:2007bf} in the space of solutions when the gravity theory has black hole solutions. By assuming that the integral is path-independent, we can define the off-shell ADT conserved charge as
\begin{align}\label{ADT charge}
   \mathcal{Q}( \xi)&\equiv \frac{1}{8 \pi  G}\int _0^1ds\int _{\Sigma }d^{D-2}x_{\mu \nu }\sqrt{-g}\mathcal{Q}_{\mathrm{ADT}}^{\mu \nu}    \nonumber  \\&=\frac{1}{16 \pi  G}\int _{\Sigma }d^{D-2}x_{\mu \nu }\left[  \Delta K^{\mu \nu } (\xi)-2\xi ^{[\mu }\int _0^1ds\,\,\tilde{\Theta }^{\nu ]}(\Phi ,s\mathcal{M}) \right] ,
\end{align}
where
\begin{equation}\label{key}
   \Delta K^{\mu \nu }=K_{s=1}^{\mu \nu }-K_{s=0}^{\mu \nu }
\end{equation}
is the finite difference of Noether potential between the given solution and the background solution, $s$ is the path parameter $(s\in [0,1])$, and $\mathcal{M}$'s are the black hole solution parameters. Eq.\eqref{ADT charge} can be used to compute quasi-local conserved charges including mass, angular momentum and entropy for a given black hole solution. 

\section{Off-shell ADT potentials in general models}
\subsection{Einstein-Hilbert theory in Palatini formalism}

In this section we will give the off-shell ADT potential of Einstein-Hilbert theory in Palatini formalism with tortion and non-metricity. The Einstein-Hilbert action with a cosmological constant in D-dimension is given by 
\begin{equation}\label{key}
%S_{EH}[g_{\mu \nu},\Gamma^{\lambda}{}_{\alpha \beta}]
I[g,\Gamma]=\frac{1}{16 \pi G} \int d^Dx \sqrt{-g} (R-2\Lambda).
\end{equation}
From Eq.\eqref{Lvariation}, the variations of the Einstein-Hilbert action give the Euler-Lagrange expressions and the surface term as
\begin{align} \label{eq43}
   \mathcal{E}^{(\mu \nu)}=&R^{(\mu \nu)}-\frac{1}{2} R g^{\mu \nu}+\Lambda g^{\mu \nu} , \\
   P_{\lambda}{}^{\nu \mu}=&-\frac{\nabla_{\lambda}(\sqrt{-g}g^{\nu \mu})}{\sqrt{-g}}+\frac{\nabla_{\sigma}(\sqrt{-g}g^{\nu \sigma})\delta^{\mu}{}_{\lambda}}{\sqrt{-g}}+(T_{\lambda}g^{\nu \mu}-T^{\nu}\delta^\mu{}_{\lambda}+g^{\nu \sigma}T^{\mu}{}_{\sigma \lambda}), \\
   \tilde \Theta^{\lambda}=&\sqrt{-g}g^{\mu \nu} \delta \Gamma^{\lambda}{}_{\mu \nu}-\sqrt{-g}g^{\nu \lambda} \delta^{\mu}{}_{\sigma} \delta \Gamma^{\sigma}{}_{\mu \nu} \label{EH surface term}.
\end{align}
The diffeomorphism variation of Lagrangian leads to 
\begin{align} \label{eq46}
   \tilde \Theta^{\lambda}(\delta_{\xi} \Phi, \Phi)=&\sqrt{-g}g^{\mu \nu} \delta_{\xi} \Gamma^{\lambda}{}_{\mu \nu}-\sqrt{-g}g^{\nu \lambda} \delta^{\mu}{}_{\sigma} \delta_{\xi} \Gamma^{\sigma}{}_{\mu \nu}  \nonumber \\
   =&\partial_{\nu}(\sqrt{-g} P_{\sigma}{}^{\nu \lambda} \xi^{\sigma})-T_{\nu}(\sqrt{-g} P_{\sigma}{}^{\nu \lambda} \xi^{\sigma})+\sqrt{-g} \Gamma^{\lambda}{}_{\rho \nu} P_{\sigma}{}^{\nu \rho} \xi^{\sigma}-\bar \nabla_{\nu}(\sqrt{-g} P_{\sigma}{}^{\nu \lambda}) \xi^{\sigma} .
\end{align}
By using the Bianchi identities \eqref{BI8}-\eqref{BI9}, we can prove that the left hand side of Eq.\eqref{eq19} vanishes in Einstein-Hilbert theory, which leads to $\mathcal{Z}^{\mu}_{\xi}=0$. From Eq.\eqref{off-shell Noether current}, we get the off-shell Noether current 
\begin{align} \label{EH Noethr curent}
   J^{\mu}_{\xi}=&2 \sqrt{-g} R^{(\mu \nu)} \xi_{\nu}+\sqrt{-g} \check R^{\mu \nu} \xi_{\nu}-\sqrt{-g} R^{\mu \nu} \xi_{\nu} \nonumber+[\nabla_{\nu} (\sqrt{-g} P_{\sigma}{}^{\mu \nu})-T_{\nu} (\sqrt{-g} P_{\sigma}{}^{\mu \nu})] \xi^{\sigma}  \nonumber \\
   &+\partial_{\nu}(2 \sqrt{-g} g^{\sigma [\mu} \bar \nabla_{\sigma} \xi^{\nu]}) .
\end{align}
By using the Bianchi identities \eqref{BI8}-\eqref{BI9} again, the first line of Eq.\eqref{EH Noethr curent} vanishes, then the off-shell Noether current and potential are
\begin{align}
   J^{\mu}_{\xi}=& \partial_{\nu}(2 \sqrt{-g} g^{\sigma [\mu} \bar \nabla_{\sigma} \xi^{\nu]}) \nonumber \\
   =& \partial_{\nu}(\sqrt{-g} \mathcal{K}^{\mu \nu}_{\xi}), \label{eq48} \\
   \mathcal{K}^{\mu \nu}_{\xi}&= 2 \bar \nabla^{[\mu} \xi^{\nu]} \label{eq49}.
\end{align}

From Eqs.\eqref{off-shell ADT potential} and \eqref{EH surface term} we obtain the final expression of the off-shell ADT potential
\begin{align} \label{EH ADT potential}
   \mathcal{Q}^{\mu \nu}_{\mathrm{ADT}}=&\frac{1}{2} h \bar{\nabla}^{[\mu} \xi^{\nu]}-h^{\sigma [\mu} \bar{\nabla}_{\sigma} \xi^{\nu]}+g^{\sigma [\mu} \delta \Gamma^{\nu]}{}_{\rho \sigma} \xi^{\rho}  \nonumber \\
   &-\xi^{[\mu} \delta \Gamma^{\nu]}{}_{\alpha \beta} g^{\alpha \beta}+\xi^{[\mu} g^{\nu] \sigma} \delta \Gamma^{\rho}{}_{\rho \sigma},
\end{align}
where we have defined
\begin{gather}
   h_{\mu \nu }=\delta g_{\mu \nu },\,\,\,\,h^{\mu \nu }= g^{\mu \alpha } g^{\nu \beta }\delta g_{\alpha \beta }=-\delta g^{\mu \nu },\,\,\,\,h=g^{\mu \nu }\delta g_{\mu \nu }.
\end{gather}
From Eq.\eqref{EH ADT potential}, we see that the torsion and non-metricity tensors do not apparently present in the off-shell ADT potential, but included through the affine connection $\Gamma^{\lambda}{}_{\mu \nu}$. On the other hand, our formalism does not match with the BBC formalism in Palatini formalism of general relativity presented in \cite{Barnich:2020ciy} when the torsion are set to zero. As pointed out in \cite{Barnich:2020ciy} this difference in Palatini theory comes from the use of the generalized Killing vector.

In the standard metric formalism, it is assumed that torsion and non-metricity are absent, i.e. $\nabla_{\alpha} g_{\mu \nu}=0, \,\,T^{\lambda}{}_{\mu \nu}=-2 \Gamma^{\lambda}{}_{[\mu \nu]}=0, \,\,\bar \nabla_{\mu}=\nabla_{\mu} $, and the connection is completely fixed to Levi-Civita connection $\mathring \Gamma^{\lambda}{}_{\mu \nu}$. Then, our expressions reduce to
\begin{align}
   &\mathcal{K}^{\mu \nu}_{\xi}=2 \nabla^{[\mu} \xi^{\nu]} ,  \nonumber \\
   &\Theta^{\mu}(\delta \Phi, \Phi)=2 \nabla^{[\sigma} h^{\mu]}{}_{\sigma} , \nonumber \\
   &\mathcal{Q}^{\mu \nu}_{\mathrm{ADT}}=\frac{1}{2} h \nabla^{[\mu} \xi^{\nu]}-h^{\sigma [\mu} \nabla_{\sigma} \xi^{\nu]}-\xi^{[\mu} \nabla_{\sigma} h^{\nu] \sigma}+\xi_{\sigma} \nabla^{[\mu} h^{\nu] \sigma}+\xi^{[\mu} \nabla^{\nu]} h ,
\end{align}
which are well known results in Einstein general relativity in the metric formalism \cite{Wald:1993nt,Iyer:1994ys,Wald:1999wa}.

\subsection{General $L(g_{\mu \nu}, R^{\lambda}{}_{\nu \alpha \mu}, T^{\lambda}{}_{\alpha \beta}, Q_{\alpha \mu \nu})$ theories}

In the gravitation sector, the most general Lagrangian that one could write down (without including additional tensors constructed by the covariant derivatives of basic geometrical objects) is $L(g_{\mu \nu}, R^{\lambda}{}_{\nu \alpha \mu}, T^{\lambda}{}_{\alpha \beta}, Q_{\alpha \mu \nu})$ \cite{Iosifidis:2019jgi}. In this subsection we will derive the off-shell ADT potential in this theory. The action is given by 
\begin{equation}\label{key}
   I[g,\Gamma]=\frac{1}{16 \pi G} \int d^Dx \sqrt{-g} L(g_{\mu \nu}, R^{\lambda}{}_{\nu \alpha \mu}, T^{\lambda}{}_{\alpha \beta}, Q_{\alpha \mu \nu}).
\end{equation}
The EOMs have been given in \cite{Iosifidis:2019jgi} for this theory. In order to derive the off-shell ADT potential we write them down and give the surface term by generic variation 
\begin{align}
   \mathcal{E}^{\mu \nu}=&-\mathcal{G}^{\mu \nu}-\frac{1}{2}g^{\mu \nu} L+T_{\sigma} W^{\sigma \mu \nu}-\frac{1}{\sqrt{-g}} \nabla_{\sigma}(\sqrt{-g} W^{\sigma \mu \nu}) , \nonumber \\
   P_{\lambda}{}^{\nu \mu}=&-\frac{2}{\sqrt{-g}}\nabla_{\sigma}(\sqrt{-g} \Omega_{\lambda}{}^{\nu \sigma \mu})-\Omega_{\lambda}{}^{\nu \sigma \rho} T^{\mu}{}_{\sigma \rho}-2V_{\lambda}{}^{\mu \nu}+2W^{\mu \nu}{}_{\lambda}+2 T_{\sigma} \Omega_{\lambda}{}^{\nu \sigma \mu} ,  \nonumber \\
   \tilde \Theta^{\alpha}(\delta \Phi, \Phi)=&2 \sqrt{-g} \Omega_{\lambda}{}^{\nu \alpha \mu} \delta \Gamma^{\lambda}{}_{\mu \nu}-\sqrt{-g} W^{\alpha \mu \nu} \delta g_{\mu \nu} ,
\end{align}
where 
\begin{align}
   \mathcal{G}^{\mu \nu}\equiv& \frac{\partial L}{\partial g_{\mu \nu}} , &W^{\alpha \mu \nu}&=W^{\alpha (\mu \nu)} \equiv \frac{\partial L}{\partial Q_{\alpha \mu \nu}} , \nonumber \\
   \Omega_{\lambda}{}^{\nu \alpha \mu}=&\Omega_{\lambda}{}^{\nu [\alpha \mu]} \equiv \frac{\partial L}{\partial R^{\lambda}{}_{\nu \alpha \mu}} , &V_{\lambda}{}^{\mu \nu}&=V_{\lambda}{}^{[\mu \nu]} \equiv \frac{\partial L}{\partial T^{\lambda}{}_{\mu \nu}} .
\end{align}
The diffeomorphism variation of the Lagrangian has the form 
\begin{align} 
   \delta_{\xi} (\sqrt{-g} L)=&\left[2\nabla_{\mu} (\sqrt{-g} \mathcal{E}^{(\mu \nu)})-2 T_{\mu} (\sqrt{-g} \mathcal{E}^{(\mu \nu)})-2\sqrt{-g} \mathcal{E}^{(\mu \lambda)} N^{\nu}{}_{\mu \lambda}-\sqrt{-g} P_{\lambda}{}^{\alpha \beta} R^{\lambda}{}_{\alpha \beta}{}^{\nu} \right] \xi_{\nu}  \nonumber \\
   +&\left[\bar \nabla_{\nu}[\nabla_{\mu}(\sqrt{-g}P_{\sigma}{}^{\nu  \mu})-T_{\mu}(\sqrt{-g}P_{\sigma}{}^{\nu  \mu})]+T_{\nu}[\nabla_{\mu}(\sqrt{-g}P_{\sigma}{}^{\nu  \mu})-T_{\mu}(\sqrt{-g}P_{\sigma}{}^{\nu  \mu})]\right]\xi^{\sigma}  \nonumber \\
   +&\partial_{\mu} (\tilde \Theta^{\mu}(\delta_{\xi} \Phi, \Phi)-2\tilde S^{\mu}(\delta_{\xi} \Phi, \Phi)).
\end{align}
For a general Palatini theory of gravity, which is diffeomorphism invariant up to a boundary term, the first two lines must vanish
\begin{align}
   &\left[2\nabla_{\mu} (\sqrt{-g} \mathcal{E}^{(\mu \nu)})-2 T_{\mu} (\sqrt{-g} \mathcal{E}^{(\mu \nu)})-2\sqrt{-g} \mathcal{E}^{(\mu \lambda)} N^{\nu}{}_{\mu \lambda}-\sqrt{-g} P_{\lambda}{}^{\alpha \beta} R^{\lambda}{}_{\alpha \beta}{}^{\nu} \right] \xi_{\nu}  \nonumber \\
   &+\left[\bar \nabla_{\nu}[\nabla_{\mu}(\sqrt{-g}P_{\sigma}{}^{\nu  \mu})-T_{\mu}(\sqrt{-g}P_{\sigma}{}^{\nu  \mu})]+T_{\nu}[\nabla_{\mu}(\sqrt{-g}P_{\sigma}{}^{\nu  \mu})-T_{\mu}(\sqrt{-g}P_{\sigma}{}^{\nu  \mu})]\right]\xi^{\sigma}  \nonumber \\
   &=0 ,
\end{align}
which can be thought of as the generalized Bianchi identity in $L(g_{\mu \nu}, R^{\lambda}{}_{\nu \alpha \mu}, T^{\lambda}{}_{\alpha \beta}, Q_{\alpha \mu \nu})$ theory. From Eqs.\eqref{off-shell Noether current} and \eqref{Noether current and potentiali} we can get the off-shell Noether current and potential as
\begin{align}
J^{\mu}_{\xi}=& \partial_{\nu}[2 \sqrt{-g} (V_{\lambda}{}^{\mu \nu} \xi^{\lambda}- \Omega_{\lambda}{}^{\sigma \mu \nu} \bar \nabla_{\sigma} \xi^{\lambda})] \nonumber \\
=& \partial_{\nu}[\sqrt{-g} \mathcal{K}^{\mu \nu}_{\xi}], \\
\mathcal{K}^{\mu \nu}_{\xi}&= 2 ( V_{\lambda}{}^{\mu \nu} \xi^{\lambda}- \Omega_{\lambda}{}^{\sigma \mu \nu} \bar \nabla_{\sigma} \xi^{\lambda}) \label{Fnp}.
\end{align}
From Eq.\eqref{off-shell ADT potential}, we can get the off-shell ADT potential
\begin{align} \label{FADTp}
   \mathcal{Q}^{\mu \nu}_{\mathrm{ADT}}=&\frac{1}{2} h V_{\lambda}{}^{\mu \nu} \xi^{\lambda}+(\delta V_{\lambda}{}^{\mu \nu}) \xi^{\lambda}-\Omega_{\lambda}{}^{\sigma \mu \nu} \left(\frac{1}{2} h \bar \nabla_{\sigma} \xi^{\lambda}+\delta \Gamma^{\lambda}{}_{\rho \sigma} \xi^{\rho} \right) \nonumber \\
   -&(\delta \Omega_{\lambda}{}^{\sigma \mu \nu}) \bar \nabla_{\sigma} \xi^{\lambda}-\xi^{[\mu} (2\Omega_{\lambda}{}^{|\beta| \nu] \alpha} \delta \Gamma^{\lambda}{}_{\alpha \beta}-W^{\nu] \alpha \beta} \delta g_{\alpha \beta}) .
\end{align} 

Taking $L=R$, we compute
\begin{equation}\label{key}
   \Omega_{\lambda}{}^{\mu \alpha \nu}=\frac{\partial R}{\partial R^{\lambda}{}_{\mu \alpha \nu}}=\delta^{\beta}{}_{\gamma} g^{\kappa \rho} \frac{\partial R^{\gamma}{}_{\kappa \beta \rho}}{\partial R^{\lambda}{}_{\mu \alpha \nu}}=g^{\mu [\nu} \delta^{\alpha]}{}_{\lambda},
\end{equation}
where we have used the fact 
\begin{equation}\label{key}
   \frac{\partial R^{\gamma}{}_{\kappa \beta \rho}}{\partial R^{\lambda}{}_{\mu \alpha \nu}}=\delta^{\gamma}{}_{\lambda} \delta^{\mu}{}_{\kappa} \delta^{[\alpha}{}_{\beta} \delta^{\nu]}{}_{\rho} .
\end{equation}
And by using the fact
\begin{equation}\label{key}
   \frac{\partial R}{\partial g^{\mu \nu}}=R_{(\mu \nu)} ,
\end{equation}
the results reduce to the Einstein-Hilbert theory present in Eqs.\eqref{eq43}-\eqref{eq46} and Eqs.\eqref{eq48}-\eqref{EH ADT potential}.

\subsection{Teleparallel Palatini theory}

As a typical example, in this section we consider the teleparallel Palatini theory. The Lagrangian has the form \cite{BeltranJimenez:2018vdo}
\begin{equation}\label{key}
   \mathcal{L}_{||}=\sqrt{-g}{L}_{||}=\frac{1}{2} \sqrt{-g} \mathbb{T}+\lambda_{\alpha}{}^{\beta \mu \nu} R^{\alpha}{}_{\beta \mu \nu}+\lambda^{\alpha \mu \nu} Q_{\alpha \mu \nu} ,
\end{equation}
where $\lambda_{\alpha}{}^{\beta \mu \nu}$ and $\lambda^{\alpha \mu \nu}$ are the Lagrange multipliers, which constrain the teleparallelity and metricity for Teleparallel Equivalent of General Relativity (TEGR). The torsion scalar is defined as
\begin{equation}\label{key}
   \mathbb{T}= S_{\alpha}{}^{\mu \nu} T^{\alpha}{}_{\mu \nu} ,
\end{equation} 
where
\begin{equation}\label{superpotential}
   S_{\alpha}{}^{\mu \nu}=aT_{\alpha}{}^{\mu \nu}+bT^{[\mu}{}_{\alpha}{}^{\nu]}+c\delta_{\alpha}{}^{[\mu} T^{\nu]}
\end{equation}
is the superpotential tensor, in which $a$, $b$ and $c$ are arbitrary constants. Varying the Lagrangian with respect to the dynamic fields and the Lagrange multipliers, we get
\begin{align}
   \delta(\sqrt{-g} L_{||})=&\sqrt{-g} \mathcal{E}_{\Phi} \delta \Phi+\partial_{\mu} \tilde \Theta^{\mu}({\delta \Phi, \Phi, \lambda})  \nonumber \\
   =&-\sqrt{-g}\mathcal{E}^{\mu \nu} \delta g_{\mu \nu}+\sqrt{-g} P_{\lambda}{}^{\nu \mu} \delta \Gamma^{\lambda}{}_{\mu \nu} \nonumber \\
   &+R^{\alpha}{}_{\beta \mu \nu} \delta \lambda_{\alpha}{}^{\beta \mu \nu}+Q_{\alpha \mu \nu} \delta \lambda^{\alpha \mu \nu}+\partial_{\mu} \tilde \Theta^{\mu}({\delta \Phi, \Phi, \lambda}) .
\end{align}
In TEGR, the teleparallelity and the dynamics are constrained by Lagrange multipliers EOMs
\begin{equation}\label{LEOM}
   R^{\alpha}{}_{\beta \mu \nu}=0,\,\,\,\,\,\,\,\,\,\,\,\,Q_{\alpha \mu \nu}=0.
\end{equation}
If we employ the constraint condition \eqref{LEOM} and do not require the EOMs of the dynamic fields $\mathcal{E}_{\Phi}=0$ to be satisfied (i.e. off-shell), the Noether potential and the ADT potential can be extracted by using Eqs.\eqref{Fnp} and \eqref{FADTp}
\begin{align} 
   \mathcal{K}^{\mu \nu}_{\xi}=&2( S_{\alpha}{}^{\mu \nu} \xi^{\alpha}- \lambda_{\alpha}{}^{\sigma \mu \nu} \bar \nabla_{\sigma} \xi^{\alpha}) , \label{TEnpotential}\\
   \mathcal{Q}^{\mu \nu}_{\mathrm{ADT}}=&\frac{1}{2} h S_{\alpha}{}^{\mu \nu} \xi^{\alpha}+(\delta S_{\alpha}{}^{\mu \nu}) \xi^{\alpha}-\lambda_{\alpha}{}^{\sigma \mu \nu} \left(\frac{1}{2} h \bar \nabla_{\sigma} \xi^{\alpha}+\delta \Gamma^{\alpha}{}_{\rho \sigma} \xi^{\rho} \right) \nonumber \\
   -&(\delta \lambda_{\alpha}{}^{\sigma \mu \nu}) \bar \nabla_{\sigma} \xi^{\alpha}-\xi^{[\mu} (2\lambda_{\rho}{}^{|\beta| \nu] \alpha} \delta \Gamma^{\rho}{}_{\alpha \beta}-\lambda^{\nu] \alpha \beta} \delta g_{\alpha \beta}) \label{TEADTpotential},
\end{align}
where we have used the equalities
\begin{align}
   V_{\alpha}{}^{\mu \nu}=\frac{\partial L_{||}}{\partial T^{\alpha}{}_{\mu \nu}}=S_{\alpha}{}^{\mu \nu},\,\,\, \Omega_{\alpha}{}^{\beta \mu \nu}=\frac{\partial L_{||}}{\partial R^{\alpha}{}_{\beta \mu \nu}}=\lambda_{\alpha}{}^{\beta \mu \nu},\,\,\, W^{\alpha \mu \nu}=\frac{\partial L_{||}}{\partial Q_{\alpha \mu \nu}}=\lambda^{\alpha \mu \nu} .
\end{align}
Since the Lagrange multipliers do not affect the dynamical field equation \cite{BeltranJimenez:2018vdo,Golovnev:2017dox}, and thus will not affect the spacetime configuration and the black hole solutions, and will not affect the final conserved charges when the explicit black hole solutions are concerned, the Lagrange multipliers terms can be dropped out from the Noether and the ADT potentials, through which Eqs.\eqref{TEnpotential} and \eqref{TEADTpotential} reduce to
\begin{align} 
   \mathcal{K}^{\mu \nu}_{\xi}=&2S_{\alpha}{}^{\mu \nu} \xi^{\alpha} , \label{KTEGR} \\
   \mathcal{Q}^{\mu \nu}_{\mathrm{ADT}}=&\frac{1}{2} h S_{\alpha}{}^{\mu \nu} \xi^{\alpha}+(\delta S_{\alpha}{}^{\mu \nu}) \xi^{\alpha} .
\end{align}
These results are exactly those obtained in \cite{Emtsova:2019moq} by the direct use of the Noether’s theorem and in \cite{Hammad:2019oyb} by Wald’s CPSM in TEGR. 

\section{Conserved charges for explicit black hole solutions}

As thermodynamic systems, black holes have a characteristic temperature, an entropy and the laws of black hole thermodynamics associated with them. In Palatini theory of gravity with torsion and non-metricity, if the black hole solutions exist, we give the general expressions of the black hole conserved charges and the first law of black hole thermodynamics. 

For simplicity, we consider a stationary axial-symmetric black hole solution with torsion and non-metricity. We adopt the coordinate in which the timelike and the rotational Killing vector are $\xi_T=\partial_t$ and $\xi_R=-\partial_{\varphi}$, respectively. The horizon Killing vector is a linear combination $\xi_{\mathrm{H}}=\partial_t+\Omega_{\mathrm{H}} \partial_{\varphi}$ that satisfies $\delta_{\xi} g_{\mu \nu}=0$ and $ \delta_{\xi} \Gamma^{\lambda}{}_{\mu \nu}=0$, where $\Omega_{\mathrm{H}}$ is the horizon angular velocity. Similar to the CPSM and the BBC formalisms, we define the black hole entropy as the conserved charge associated with horizon Killing vector $\xi_{\mathrm{H}}$,
\begin{equation}\label{key}
\frac{\kappa}{2 \pi} \delta S_{\mathrm{H}}=\delta \mathcal{Q} (\xi_{\mathrm{H}})=\frac{1}{8 \pi G} \int_{\mathcal{H}} d^{D-2}x_{\mu \nu} \sqrt{-g} \mathcal{Q}^{\mu \nu}_{\mathrm{ADT}}(\xi_{\mathrm{H}}) ,
\end{equation} 
where $\kappa$ is the surface gravity defined as $\kappa=-n^{\nu} \xi^{\mu} \nabla_{\nu} \xi_{\mu}$ in the presence of torsion \cite{Dey:2017fld}, leading to $T_{\mathrm{H}}=\kappa_{\mathrm{H}}/{2 \pi}$. From the linearity of $\delta \mathcal{Q}(\xi)$ in $\xi$, we can obtain the first law
\begin{equation}\label{key}
   T_{\mathrm{H}}\delta S_{\mathrm{H}}=\delta M-\Omega_{\mathrm{H}} \delta J  ,
\end{equation}
where 
\begin{equation}\label{key}
   \delta M=\delta \mathcal{Q}(\xi_T)=\frac{1}{8 \pi G} \int_{\Sigma} d^{D-2}x_{\mu \nu} \sqrt{-g} \mathcal{Q}^{\mu \nu}_{\mathrm{ADT}}(\xi_T)
\end{equation}
and 
\begin{equation}\label{key}
   \delta J=\delta \mathcal{Q}(\xi_R)=\frac{1}{8 \pi G} \int_{\Sigma} d^{D-2}x_{\mu \nu} \sqrt{-g} \mathcal{Q}^{\mu \nu}_{\mathrm{ADT}}(\xi_R)
\end{equation}
are the mass and angular momentum variations associated with $\xi_T$ and $\xi_R$, respectively.
Similar to SPSM \cite{Hajian:2015xlp,Hajian:2016kxx,Ghodrati:2016vvf} and our previous work \cite{Ding:2019dhy}, $\Sigma$ can be an almostly arbitrary smooth co-dimensional two surface surrounding the singularity.
Our above formulation can be used to compute the conserved charges and derive the first law of black hole thermodynamics in the modified gravity theories with torsion and non-metricity (e.g. Palatini theory, MAG and TEGR etc.). 

As a simple example, we now check the total mass of Schwarzschild black hole in the teleparallel theory of gravity. In spherical coordinates the Schwarzschild metric is given by
\begin{equation}\label{Schmetric}
	ds^2=-(1-\frac{2 m}{r}) dt^2+(1-\frac{2 m}{r})^{-1} dr^2+r^2(d \theta^2+\mathrm{sin}^2\theta d \varphi^2) .
\end{equation}
It has the non-zero components of affine connection \cite{Emtsova:2019moq}
\begin{align}
	\Gamma^{t}{}_{t r}&=-\Gamma^{r}{}_{r r}=-\frac{m}{r^2}(1-\frac{2 m}{r})^{-1},&\,\,\,\,\,\,&\Gamma^{r}{}_{\theta \theta}=r(1-\frac{2 m}{r})^{1/2},  \nonumber \\
	\Gamma^{r}{}_{\varphi \varphi}&=r(1-\frac{2 m}{r})^{1/2} r\, \mathrm{sin}^2 \theta,&\,\,\,\,\,\,&\Gamma^{\theta}{}_{r \theta}=\Gamma^{\varphi}{}_{r \varphi}=-\frac{1}{r}(1-\frac{2 m}{r})^{-1/2},  \nonumber \\
	\Gamma^{\theta}{}_{\theta r}&=\Gamma^{\varphi}{}_{\varphi r}=-\frac{1}{r},&\,\,\,\,\,\,&\Gamma^{\theta}{}_{\varphi \theta}=\mathrm{sin}\theta \,\mathrm{cos} \theta,\,\,\,\,\,\Gamma^{\varphi}{}_{\theta \varphi}=\Gamma^{\varphi}{}_{\varphi \theta}=-\mathrm{cot}\theta ,
\end{align}
From the first formula of \eqref{gobjects}, we get the non-zero independent components of torsion tensor
\begin{align}
	T^{t}{}_{t r}=\frac{m}{r^2}(1-\frac{2 m}{r})^{-1},\,\,\,\,\,\,\,\,\,\,\,\,T^{\theta}{}_{r \theta}=T^{\varphi}{}_{r \varphi}=-\frac{1}{r}(1-(1-\frac{2 m}{r})^{-1/2}) .
\end{align}
From Eq.\eqref{superpotential}, the non-zero independent components of the superpotential are given by
\begin{equation}\label{key}
	S_{t}{}^{t r}=\frac{1}{r}(1-(1-\frac{2 m}{r})^{1/2})-\frac{2 m}{r^2},\,\,\,\,\,\,\,S_{\theta}{}^{r \theta}=S_{\varphi}{}^{r \varphi}=-\frac{1}{2r}(1-(1-\frac{2 m}{r})^{1/2})+\frac{m}{2r^2}.
\end{equation}
By using Eqs.\eqref{ADT charge}, \eqref{KTEGR}, choosing the one parameter path by substituting $m$ by $s m$ in the Schwarzschild metric \eqref{Schmetric}, choosing the timelike Killing vector $\xi^{\mu}=(-1,0,0,0)$, taking the integral surface $\Sigma: t=const=r$ for convenience, and taking the limit $r\to \infty$, we get the total mass of Schwarzschild black hole as
\begin{align}
	M&=\frac{1}{8 \pi  G} \lim_{r\to \infty} \int _0^1ds\int _{\Sigma }d^{D-2}x_{\mu \nu }\sqrt{-g}\mathcal{Q}_{\mathrm{ADT}}^{\mu \nu}  \nonumber \\&=\frac{1}{16 \pi  G} \lim_{r\to \infty} \int _{\Sigma }d^2x_{t r }2\sqrt{-g} S_{t}{}^{t r} \xi^{t}  \nonumber \\
	&=\frac{m}{G} .
\end{align}
Therefore, we get the correct conserved charge (mass) of Schwarzschild black hole in teleparallel theory of gravity by our generalized off-shell ADT conserved charge formalism.

\section{Conclusions}

In this paper we have generalized the off-shell ADT conserved charge formalism to Palatini theory of gravity with torsion and non-metricity. For a general Palatini theory of gravity, which is diffeomorphism invariant up to a boundary term, we obtained the most general expression of off-shell ADT potential. As explicit examples, we derived the off-shell ADT potentials for Einstein-Hilbert action, the most general $L(g_{\mu \nu}, R^{\lambda}{}_{\nu \alpha \mu}, T^{\lambda}{}_{\alpha \beta}, Q_{\alpha \mu \nu})$ theories and the teleparallel Palatini gravity. 

In the non-Riemannian geometrical framework, if the black hole solutions exist for a gravity theory, we defined the black hole entropy as a conserved charge, and the first law of black hole thermodynamics was derived. It is easy to generalize our off-shell formalism to arbitrary diffeomorphism invariant theory of gravity with torsion and non-metricity, even including the matter field (i.e. the MAG) by using the ‘exact symmetries’.

\end{document}